\documentclass[epsfig]{article}
\usepackage{amssymb}
\usepackage{graphicx}
\usepackage{amsmath}
\usepackage{graphicx,pstricks,vmargin,subfigure,float,amsmath,epsfig}

\input epsf.sty
\input{tcilatex}
\begin{document}

\title{{\LARGE Chaplygin gas Braneworld Inflation according to} {\LARGE %
WMAP7 }}
\author{R.Zarrouki$^{1}$ and M.Bennai$^{1,2}$\thanks{%
Corresponding authors: mdbennai@yahoo.fr, m.bennai@univh2m.ac.ma} \\
$^{1}${\small Laboratoire de Physique de la Mati\`{e}re Condens\'{e}e
(URAC10),}\\
{\small \ Facult\'{e} des Sciences Ben M'sik, B.P. 7955, Universit\'{e}
Hassan II-Mohammedia, Casablanca, Maroc. }\\
{\small \ }$^{2}${\small Groupement National de Physique des Hautes
Energies, Focal point, LabUFR-PHE, Rabat, Maroc.}}
\maketitle

\begin{abstract}
We consider a Chaplygin gas model with an exponential potential in framework
of Braneworld inflation. We apply the slow-roll approximation in the high
energy limit to derive various inflationary spectrum perturbation
parameters. We show that the inflation observables depend only on the 
e-folding number $N$ and the final value of the slow roll parameter $%
\varepsilon _{end}$. Whereas for small running of the scalar spectral index $%
\frac{dn_{s}}{d\ln k}$, the inflation observables are in good agreement with
recent WMAP7 data.


\end{abstract}

\tableofcontents

\newpage

\section{Introduction}

In the last few years, Braneworld inflation\cite{Braneworld} became a
central paradigm of cosmology. In this context, \emph{Randall-Sundrum} type $%
2$ model\cite{RSII-1} has attracted a lot of interests. To describe early
inflation and recent acceleration of the universe, various Braneworld
cosmological models have been proposed\cite{models} and deeply studied\cite%
{BSCZ}. Recently, a new Chaplygin gas model\cite{Chaplygin} was introduced
to describe dark matter and dark energy recently discovered\cite{DARK M E}.
Chaplygin gas model is a particular kind of matter, characterized by an
exotic equation of state and was applied to study various cosmological
models, such as chaotic canonical scalar field inflation\cite{Herrera1} and
exponential tachyonic inflation on the brane\cite{Herrera}. More recently, a
generalized version of the original Chaplygin gas model has been proposed
and studied in the context of late time acceleration of the Uuniverse\cite%
{GCG} and in the context of Braneworld model\cite{Bouhmadi1}. Note that
Chaplygin gas has also recently been used to describe early inflationary
universe\cite{Bouhmadi2}. In the Chaplygin gas inspired inflation model, the
scalar field is usually the standard inflaton field, where the energy
density can be extrapolated to obtain a successful inflationary period with
a Chaplygin gas model\cite{Bertolami}. The motivation for introducing
Chaplygin gas Braneworld models is the increasing interest in studing dark
matter and dark energy in higher-dimensional cosmological models. We signal
in this context, that Chaplygin gas model is considered now as a viable
alternative model that can provide an accelerated expansion of the early
universe. The Chaplygin gas emerges as an effective fluid of a generalized
d-brane in space time, where the action can be written as a generalized
Born-Infeld action\cite{Bento}.

In this work, our aim is to quantify the modifications of the Chaplygin gas
inspired inflation in the Braneworld scenario. Recall that the generalized
Chaplygin gas is defined by an exotic equation of state of the form\cite%
{Bento}%
\begin{equation}
p_{_{ch}}=-\frac{A}{\rho _{_{ch}}^{\alpha }},
\end{equation}%
where $\rho _{ch}$ and $p_{ch}$ are the energy density and pressure of the
generalized Chaplygin gas, respectively. $\alpha $ is a constant satisfaying 
$0\prec \alpha \preceq 1$, and $A$ is a positive constant. Fom the matter
conservation equation, one can obtain the following generalized Chaplygin
gas energy density expression 
\begin{equation}
\rho _{_{ch}}=\left[ A+\left( \rho _{_{ch_{0}}}^{\alpha +1}-A\right) \left( 
\frac{a_{_{0}}}{a}\right) ^{3\left( \alpha +1\right) }\right] ^{\frac{1}{%
\alpha +1}},
\end{equation}%
where $a_{0}$ and $\rho _{_{ch_{0}}}$ are the present day values of the
scale factor and the generalized Chaplygin gas energy density, respectively.
Note that the original Chaplygin gas model corresponds to $\alpha =1$. In
this case, its was shown that the generalized Chaplygin gas model can
describe late time acceleration of the Universe either for small values of
the parameter $\alpha $ or for very large ones\cite{Oliver}. In order to
describe early inflationary universe, we can use the following extrapolation%
\cite{Bertolami}%
\begin{equation}
\rho _{ch}=\left[ A+\rho _{m}^{\left( \alpha +1\right) }\right] ^{\frac{1}{%
\alpha +1}}\longrightarrow \rho _{ch}=\left[ A+\rho _{\phi }^{\left( \alpha
+1\right) }\right] ^{\frac{1}{\alpha +1}}.
\end{equation}%
In this work, we first start in section $2$, by recalling the foundations of
Chaplygin gas Braneworld inflation and in particular, the modified \emph{%
Friedmann} equation and various perturbation spectrum parameters are given.
In the section $3,$ we present our results for an exponential potential, and
show that for some values of the number of e-folding  $N$, the inflation
parameters are in good agreement with recent WMAP7 data. The conclusions are
given in the last section.

\section{Chaplygin gas Braneworld inflation}

\subsection{Slow-Roll approximation}

We start this section by recalling briefly some fundations of \emph{%
Randall-Sundrum} type $2$ Braneworld model. In this scenario, our universe
is considered as a $3$-brane embedded in $five$-dimensional anti-de Sitter
space-time ($AdS5$) where gravitation can propagate through a supplementary
dimension. One of the most relevant consequences of this model is the
modification of the \emph{Friedmann }equation for energy density of the
order of the brane tension or higher. \bigskip In the case where the matter
in the brane is domintad by the generalized Chaplygin gas (GCG), the
gravitationnal \emph{Einstein} equations lead to the modified \emph{Friedmann%
} equation on the brane \cite{Herrera1,Bertolami}:%
\begin{equation}
H^{2}=\frac{8\pi }{3m_{p}^{2}}\left( A+\rho _{\phi }^{\left( \alpha
+1\right) }\right) ^{\left( \frac{1}{\alpha +1}\right) }\left( 1+\frac{%
(A+\rho _{\phi }^{\left( \alpha +1\right) })^{\frac{1}{\alpha +1}}}{2\lambda 
}\right) ,
\end{equation}%
where $\rho _{\phi }=\frac{\overset{\cdot }{\phi }^{2}}{2}+V\left( \phi
\right) $ and $V\left( \phi \right) $ is the scalar field potential
responsible of inflation, $\lambda $ is the brane tension and $m_{p}$ is the 
\emph{Planck} mass.

One has also a second inflationary relation given by the \emph{Klein-Gordon}
Eq. governing the dynamic of the scalar field $\phi :$%
\begin{equation}
\ddot{\phi}+3H\dot{\phi}+V^{\prime }=0,
\end{equation}%
where $\dot{\phi}=\frac{\partial \phi }{\partial t}$, $\ddot{\phi}=\frac{%
\partial ^{2}\phi }{\partial t^{2}}$ and $V^{\prime }=\frac{\partial V}{%
\partial \phi }.$

In the following, we will consider the case where $\alpha =1.$ During
inflation, the energy density associated to the scalar field is comparable
to the scalar potential, i.e. $\rho _{\phi }\simeq V\left( \phi \right) $.
Here we shall introduce the well known slow-roll conditions, i.e. $\dot{\phi}%
^{2}\ll $ $V\left( \phi \right) $ and $\ddot{\phi}\ll $ $H\dot{\phi}$. Note
that, perturbations on the brane are, generally, coupled to the bulk metric
perturbations. This make the situation more complicated. However, on large
scales on the brane, the density perturbations decouple from the bulk metric
perturbations and the extreme slow-roll limit (de Sitter) become applicable
for RS-type II model\cite{Maartens}. In this case, the \emph{Friedmann}
equation reduces to%
\begin{equation}
H^{2}=\frac{8\pi }{3m_{p}^{2}}\sqrt{A+V^{2}}\left( 1+\frac{(A+V^{2})^{\frac{1%
}{2}}}{2\lambda }\right) ,
\end{equation}%
In the some context, one can also consider the slow-roll parameters to study
the spectrum of the perturbation\cite{Maartens}. The two first parameters
are given for GCG model by%
\begin{eqnarray}
\varepsilon  &=&\frac{m_{p}^{2}}{16\pi }\frac{VV^{\prime ^{2}}}{(A+V^{2})^{%
\frac{3}{2}}}\frac{\left( 1+\frac{(A+V^{2})^{\frac{1}{2}}}{\lambda }\right) 
}{\left( 1+\frac{(A+V^{2})^{\frac{1}{2}}}{2\lambda }\right) ^{2}}, \\
\eta  &=&\frac{m_{p}^{2}}{8\pi }\frac{V^{\prime \prime }}{(A+V^{2})^{\frac{1%
}{2}}\left( 1+\frac{(A+V^{2})^{\frac{1}{2}}}{2\lambda }\right) },
\end{eqnarray}%
where $V^{\prime }=\frac{\partial V}{\partial \phi }$ and $V^{\prime \prime
}=\frac{\partial ^{2}V}{\partial \phi ^{2}}$. Note that during inflation,
the conditions$\ \epsilon \ll 1$ and $\mid \eta \mid \ll 1$ are satisfied.

On the other hand, one can derive the number of e-folding as 
\begin{equation}
N=-\frac{8\pi }{m_{p}^{2}}\int_{V_{\ast }}^{V_{end}}\frac{\sqrt{A+V^{2}}}{%
V^{\prime ^{2}}}\left( 1+\frac{(A+V^{2})^{\frac{1}{2}}}{2\lambda }\right) dV,
\end{equation}%
where $V_{\ast }$ and $V_{end}$ are the values of the potentials at the
horizon exit and the end of inflation, respectively.

\subsection{Chaplygin gas perturbation spectrum}

The spectrum of the inflationary perturbations is produced by quantum
fluctuations of the scalar field around its homogeneous background values.
Thus, the power spectrum of the curvature perturbations $P_{R}\left(
k\right) $ is given in the slow-roll approximation by the following
expression\cite{Maartens}%
\begin{eqnarray}
P_{R}\left( k\right)  &=&\left( \frac{H^{2}}{2\pi \overset{\cdot }{\phi }}%
\right) ^{2} \\
&=&\frac{128\pi }{3m_{p}^{6}}\frac{(A+V^{2})^{\frac{3}{2}}}{V^{\prime ^{2}}}%
\left( 1+\frac{(A+V^{2})^{\frac{1}{2}}}{2\lambda }\right) ^{3},
\end{eqnarray}%
Note that, the power spectrum of the curvature perturbations $P_{R}\left(
k\right) $ allows us to define the scalar spectral index $n_{s}$ as\cite%
{Lyth} 
\begin{eqnarray}
n_{s}-1 &=&\frac{d\ln P_{R}\left( k\right) }{d\ln k} \\
&=&\frac{m_{p}^{2}}{8\pi (A+V^{2})^{\frac{1}{2}}\left( 1+\frac{(A+V^{2})^{%
\frac{1}{2}}}{2\lambda }\right) }\times \left( -3\frac{VV^{\prime ^{2}}}{%
(A+V^{2})}\frac{\left( 1+\frac{(A+V^{2})^{\frac{1}{2}}}{\lambda }\right) }{%
\left( 1+\frac{(A+V^{2})^{\frac{1}{2}}}{2\lambda }\right) }+2V^{\prime
\prime }\right) .
\end{eqnarray}%
Another important inflationary spectrum parameter is the amplitude of the
tensorial perturbations $P_{g}\left( k\right) ,$ describing the primordial \
gravitational wave perturbations produced by a period of extreme slow-roll
inflation, which is defined by\cite{Langlois}%
\begin{eqnarray}
P_{g}\left( k\right)  &=&\frac{64\pi }{m_{p}^{2}}\left( \frac{H}{2\pi }%
\right) ^{2}F^{2}\left( x\right)  \\
&=&\frac{128}{3m_{p}^{4}}\sqrt{A+V^{2}}\left( 1+\frac{(A+V^{2})^{\frac{1}{2}}%
}{2\lambda }\right) F^{2}\left( x\right) 
\end{eqnarray}%
where $x=Hm_{p}\sqrt{\frac{3}{4\pi \lambda }}$ and $F^{2}\left( x\right)
=\left( \sqrt{1+x^{2}}-x^{2}\sinh ^{-1}\left( \frac{1}{x}\right) \right)
^{-1}$. Note that in the high-energy limit ($\sqrt{A+V^{2}}\gg \lambda $),
we have $F^{2}\left( x\right) \approx \frac{3}{2}x=\frac{3}{2}\frac{\sqrt{%
A+V^{2}}}{\lambda }$ and in the low-energy limit ($\sqrt{A+V^{2}}\ll \lambda 
$), $F^{2}\left( x\right) \approx 1.$ Therefore, we recover the standard 4D
results at the low-energy limit as expected.

The ratio of tensor to scalar perturbations and the running of the scalar
index are presented respectively by%
\begin{eqnarray}
r\left( k\right)  &=&\left( \frac{P_{g}\left( k\right) }{P_{R}\left(
k\right) }\right) _{\mid _{k=k_{\ast }}} \\
&=&\left( \frac{m_{p}^{2}}{\pi }\frac{V^{\prime ^{2}}F^{2}\left( x\right) }{%
\left( A+V^{2}\right) \left( 1+\frac{(A+V^{2})^{\frac{1}{2}}}{2\lambda }%
\right) ^{2}}\right) _{\mid _{k=k_{\ast }}}
\end{eqnarray}%
Here, $k_{\ast }$ correspond to the case $k=Ha$; the value when the universe
scale crosses the \emph{Hubble} horizon during inflation%
\begin{equation}
\frac{dn_{s}}{d\ln k}=\frac{m_{p}^{2}}{4\pi }\frac{V^{\prime }}{\sqrt{A+V^{2}%
}}\frac{1}{\left( 1+\frac{(A+V^{2})^{\frac{1}{2}}}{2\lambda }\right) }\left(
3\frac{\partial \varepsilon }{\partial \phi }-\frac{\partial \eta }{\partial
\phi }\right) 
\end{equation}%
In the next section, we consider an exponential potential to evaluate all
the previous spectrum parameters describing Branewold inflation.

\section{Exponential potential in high energy limit}

The exponential potential was studied in various occasions, for example the
authors of ref.\cite{Copeland} have shown that inflation becomes possible in
Braneworld model for a class of potentials ordinarily too steep to support
inflation. This type of potential was also used for tachyonic inflation on
the brane\cite{Sami} and recently for tachyonic Chaplygin gas inflation on
the brane\cite{Herrera}. Here we consider an exponential potential of the
form 
\begin{equation}
V\left( \phi \right) =V_{0}\exp \left( -\frac{\beta }{m_{p}}\phi \right)
\end{equation}%
where $\beta $ and $V_{0}$ are constant.

In the present case, we apply this potential to derive and study the
behaviour of various spectrum paramaters in the Braneworl Chaplygin gas
scenario. In the high-energy limit; i.e. $\sqrt{A+V^{2}}\gg \lambda ,$ all
the inflationary parameters will be simplified. In this case, the slow-roll
parameters Eqs.(7,8) become 
\begin{eqnarray}
\varepsilon  &=&\frac{\beta ^{2}}{4\pi }\frac{\lambda V^{3}}{(A+V^{2})^{2}},
\\
\eta  &=&\frac{\beta ^{2}}{4\pi }\frac{\lambda V}{(A+V^{2})}.
\end{eqnarray}%
This give new expressions for the scalar spectral index%
\begin{equation}
n_{s}-1=\frac{\beta ^{2}}{4\pi }\frac{\lambda V}{(A+V^{2})}\left( -6\frac{%
V^{2}}{(A+V^{2})}+2\right) ,
\end{equation}%
and for the ratio of tensor to scalar perturbations%
\begin{equation}
r=\frac{6\beta ^{2}}{\pi }\frac{\lambda V^{^{2}}}{(A+V^{2})^{\frac{3}{2}}}.
\end{equation}%
On the other hand, the running of the scalar index for Chaplygin gas model
becomes 
\begin{equation}
\frac{dn_{s}}{d\ln k}=\frac{\beta ^{4}}{8\pi ^{2}}\frac{\lambda ^{2}V^{2}}{%
(A+V^{2})^{4}}\left( -A^{2}+9AV^{2}-2V^{4}\right) .
\end{equation}%
Finally, from Eq.(9), one can deduce the following expression for the
e-folding number in the high energy limit 
\begin{equation}
N=-\frac{4\pi }{\lambda \beta ^{2}}\left( V_{end}-V_{\ast }+A\left( \frac{1}{%
V_{\ast }}-\frac{1}{V_{end}}\right) \right) 
\end{equation}%
By combining Eqs.(20) and (21), we obtain 
\begin{equation}
\eta =\frac{(A+V^{2})}{V^{2}}\varepsilon .
\end{equation}%
The last equation shows that $\eta \succ \varepsilon $, then inflation can
end at $\eta =1$. Thus, the Eq.(26) implies that 
\begin{equation}
A=V_{end}^{2}\left( \frac{1-\varepsilon _{end}}{\varepsilon _{end}}\right) 
\end{equation}%
We see then that, the parameter A can be expressed in terms of $V_{end}$ and 
$\varepsilon _{end}.$ Notice as well, that since $A\succ 0$ then $%
\varepsilon _{end}\prec 1$.

Inserting theexpression of $A$ (Eq.(27)) in Eq.(21), we obtain%
\begin{equation}
V_{end}=\frac{\lambda \beta ^{2}}{4\pi }\varepsilon _{end}.
\end{equation}%
According to Eq.(25), we find%
\begin{equation}
V_{\ast }=\mu V_{end},
\end{equation}%
where \ \ 
\begin{equation}
\mu =\frac{\frac{N}{\varepsilon _{end}}+\varepsilon _{end}+\sqrt{\left( 
\frac{N}{\varepsilon _{end}}+\varepsilon _{end}\right) ^{2}+4\left( \frac{%
1-\varepsilon _{end}}{\varepsilon _{end}}\right) }}{2}.
\end{equation}%
We signal that, in the limit where $A\longrightarrow 0$, we obtain 
\begin{eqnarray}
\varepsilon _{end} &=&1 \\
\mu  &=&N+1
\end{eqnarray}%
Consequently, we can find all the inflationary parameters in terms of $%
V_{\ast }$ using Eqs.(28,29)%
\begin{equation}
n_{s}-1=\frac{\mu }{\varepsilon _{end}\left( \mu ^{2}+\frac{1-\varepsilon
_{end}}{\varepsilon _{end}}\right) }\left( \frac{-6\mu ^{2}}{\left( \mu ^{2}+%
\frac{1-\varepsilon _{end}}{\varepsilon _{end}}\right) }+2\right) .
\end{equation}%
The equation (33) shows that the scalar spectral index $n_{s}$ depends only
on $N$ and $\varepsilon _{end}$. In the limit where $A\longrightarrow 0$, we
get%
\begin{equation}
n_{s}-1=-\frac{4}{N+1}.
\end{equation}%
This result was first pointed out in ref.\cite{Copeland}.

Note that for $N=50,$ we have $n_{s}\simeq 0.9215$ whereas for $N=55$, $%
n_{s}\simeq 0.9285$ and for $N=60,$ $n_{s}\simeq 0.9344.$ These values are
outside the range given by WMAP7 observation\cite{WMAP7}, since%
\begin{equation}
0.963\preceq n_{s}\preceq 1.002\text{\ \ }(95\%CL)\text{ }
\end{equation}%
Thus, generally the e-folding number must be very large.

On the other hand, in the Chaplygin gas model, we have also derived the
ratio of tensor to scalar perturbations, which is given by%
\begin{equation}
r=\frac{24\mu ^{2}}{\varepsilon _{end}\left( \mu ^{2}+\frac{1-\varepsilon
_{end}}{\varepsilon _{end}}\right) ^{\frac{3}{2}}}.
\end{equation}%
Similarly, in the limit where $A\longrightarrow 0$, $r$ becomes%
\begin{equation}
r=\frac{24}{N+1}.
\end{equation}%
As for $n_{s}$, we recover here the results of ref.\cite{Copeland} . The
Eq.(37) denotes that the number of e-folding $N$ must satisfy the inequality 
$N\succ 65$ in order for $r$ to be consistent with WMAP7 since\cite{WMAP7}%
\begin{equation}
r<0.36\text{ \ \ }(95\%CL)
\end{equation}%
In the original Chaplygin exponential Braneworld inflation model ($\alpha =1)
$, the running of the scalar index is%
\begin{equation}
\frac{dn_{s}}{d\ln k}=\frac{2\mu ^{2}}{\varepsilon _{end}^{2}\left( \mu ^{2}+%
\frac{1-\varepsilon _{end}}{\varepsilon _{end}}\right) ^{4}}\left( -2\mu
^{4}+9\mu ^{2}\left( \frac{1-\varepsilon _{end}}{\varepsilon _{end}}\right)
-\left( \frac{1-\varepsilon _{end}}{\varepsilon _{end}}\right) ^{2}\right) .
\end{equation}%
Similarly to the previous case, in the limit where $A\longrightarrow 0$, we
get%
\begin{equation}
\frac{dn_{s}}{d\ln k}=-\frac{4}{\left( N+1\right) ^{2}}.
\end{equation}%
This equation shows that the running of the scalar index is small and
negative for any value of e-folding number $N,$ which correspond to
sufficient inflation.

To confront simutaniously the observables $n_{s}$, $r$ and $\frac{dn_{s}}{%
dlnk}$ with observations, we study the relative variation of these
parameters. 
\begin{figure}[tbp]
\begin{center}
\includegraphics[height=7.5cm,width=10.5cm]{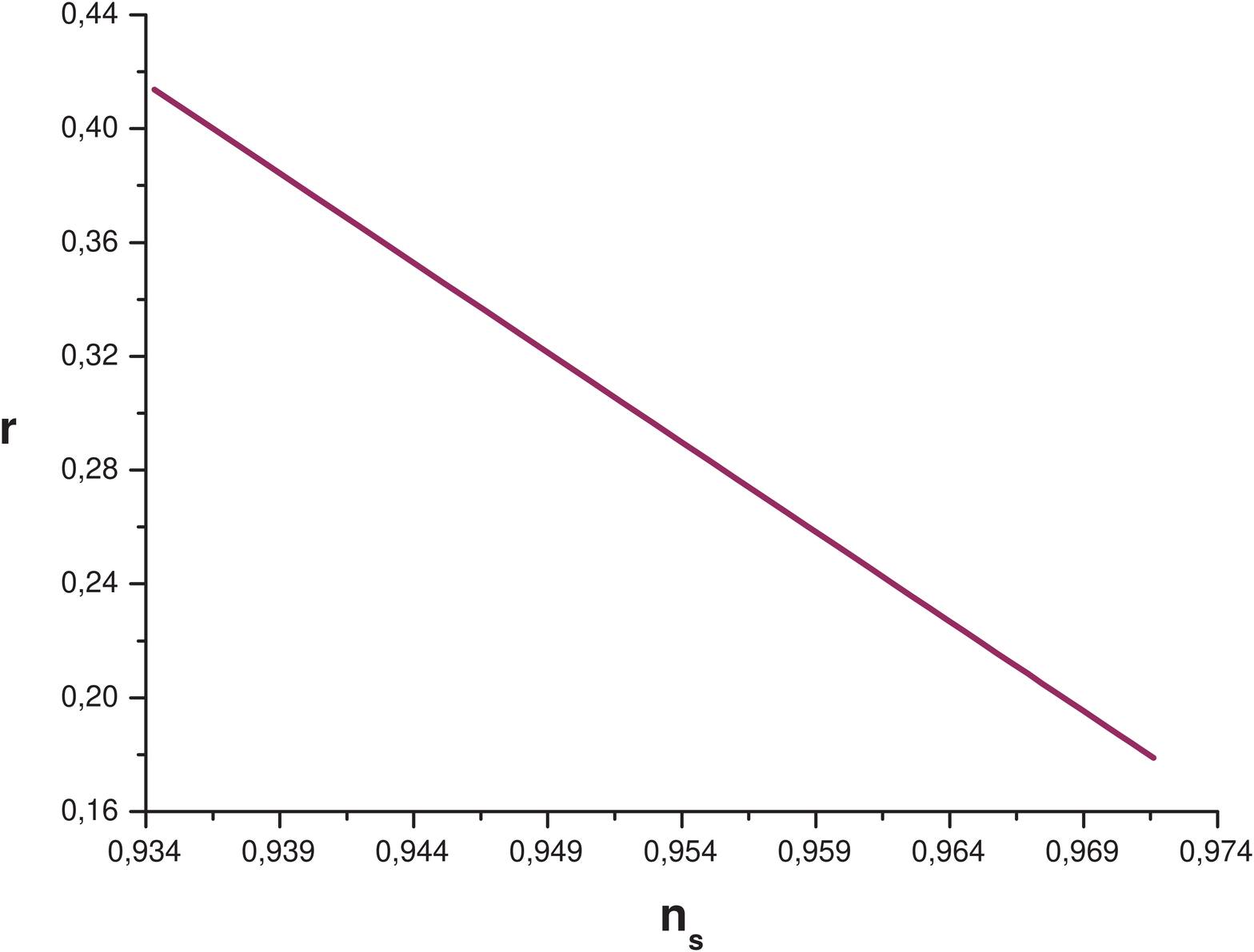}
\end{center}
\caption{$r$ vs $n_{s}$ for exponential potential in Chaplygin gas
Braneworld inflation}
\end{figure}
Fig.1, shows that the paramater $r$ behaves as a decreasing function with
respect to $n_{s}$. We observe that a large domain of variations of $r$ is
consistent with WMAP7 data. We note also that, the scalar spectral index
values are in good agreement with the observations for large $N$. 
\begin{figure}[tbp]
\begin{center}
\includegraphics[height=10.5cm,width=18.5cm]{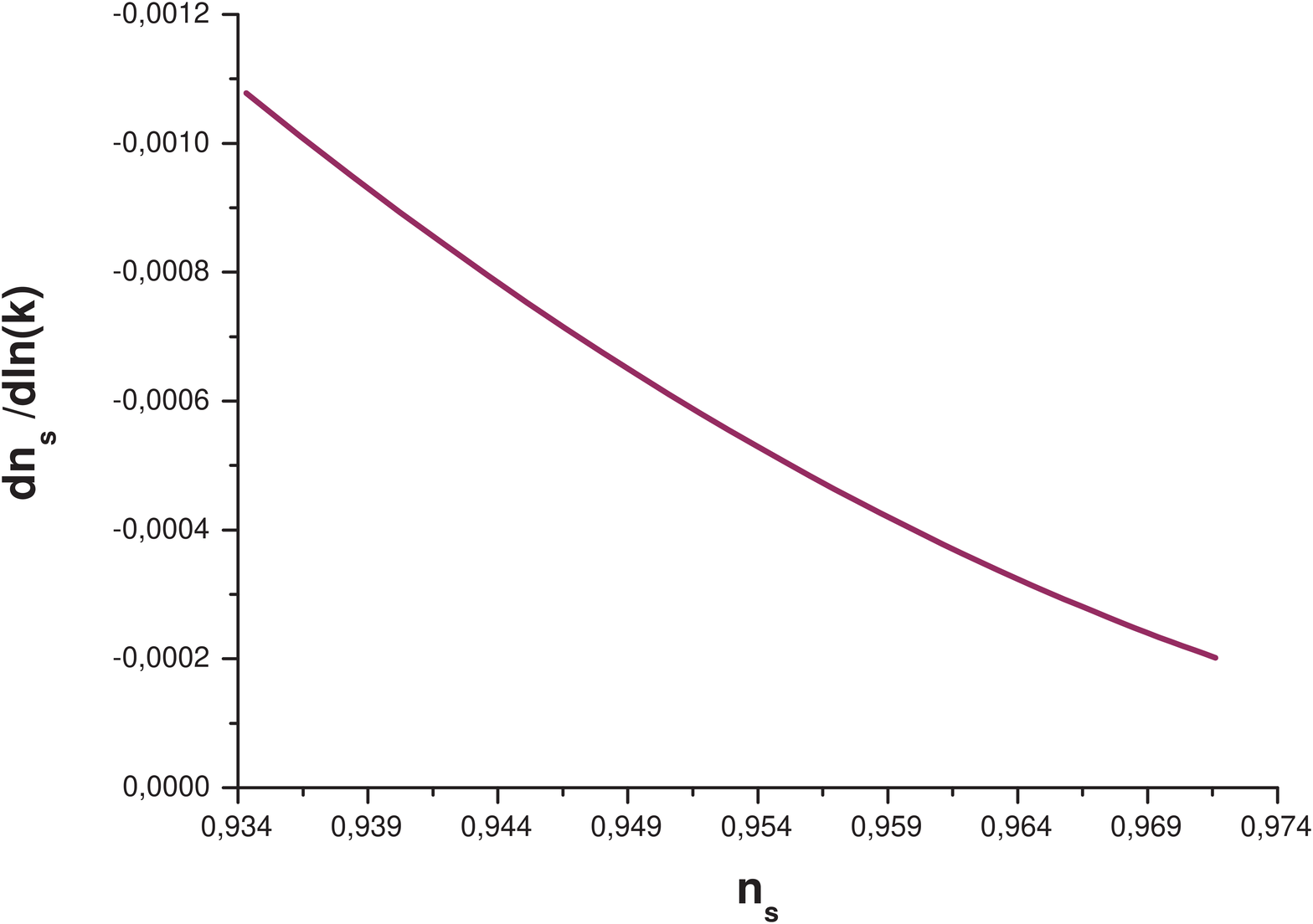}
\end{center}
\caption{$\frac{dns}{dlnk}$ vs $n_{s}$ for exponential potential in
Chaplygin gas Braneworld inflation}
\end{figure}
In Fig.2, we plot the running of the scalar index $\frac{dn_{s}}{d\ln k}$
versus the scalar spectral index $n_{s}.$ We show that the range, in which
the scalar spectral index $n_{s}$ is consistent with the recent data,
corresponding to negligible running. 
\begin{figure}[tbp]
\begin{center}
\includegraphics[height=12.5cm,width=18.5cm]{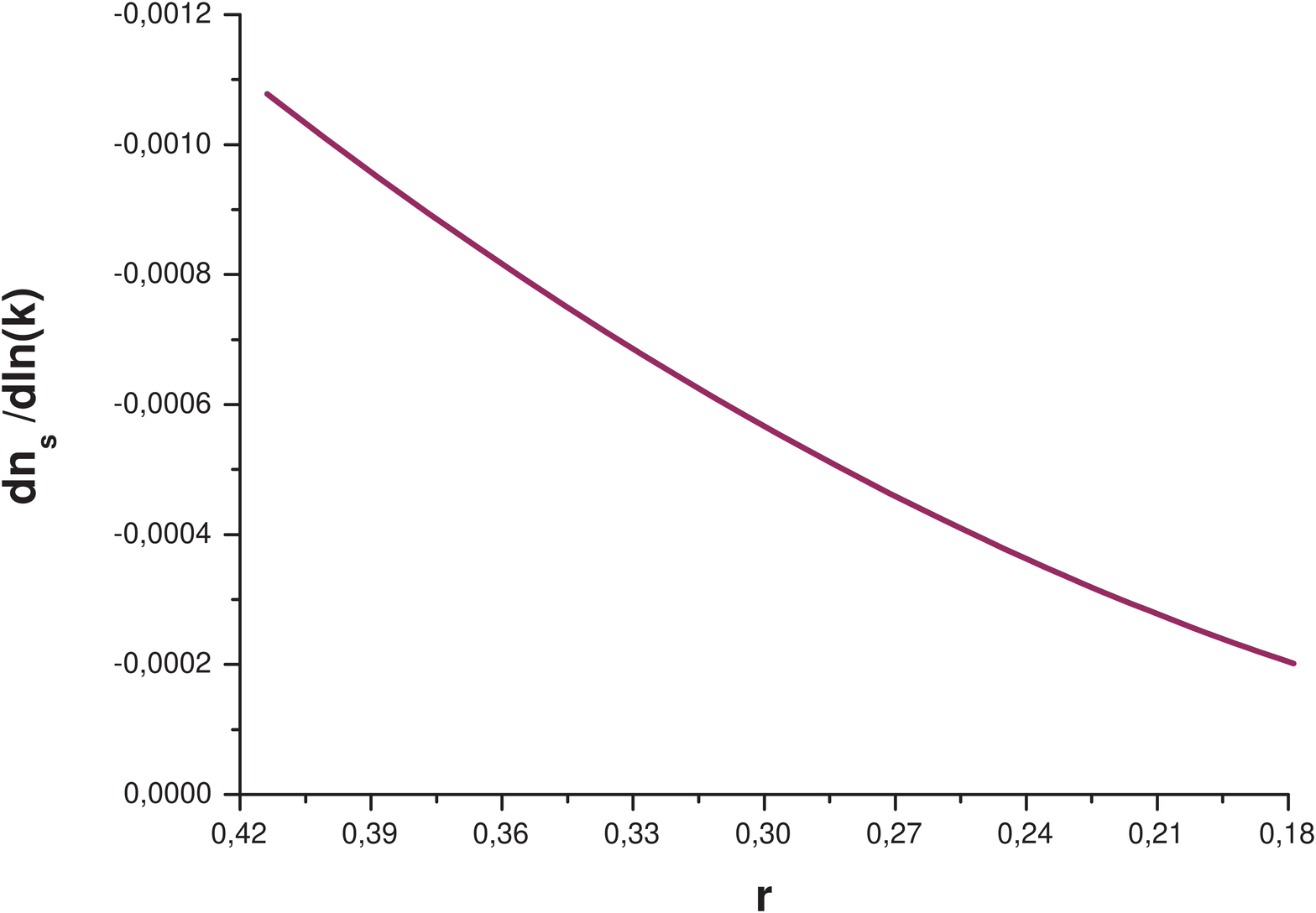}
\end{center}
\caption{$\frac{dns}{dlnk}$ vs r for exponential potential in Chaplygin gas
Braneworld inflation}
\end{figure}
In the last figure, we show that $\frac{dn_{s}}{d\ln k}$ is a decreasing
function according to $r$. We observe in this case that, for any values of $r
$ allowed by the observations (Eq.38), the running of the scalar index $%
\frac{dn_{s}}{d\ln k}$ is small and negative. This result allows us to
consider a negligible running which justifies our choice of observable
values for the inflationary parameters.

\section{Conclusion}

In this paper, we have studied a Chaplygin gas model in the framework of
Braneworld inflation using an exponential potential. We have adopted here
the slow roll approximation in the high-energy limit to derive all
inflationary spectrum perturbation parameters in the particular case where $%
\alpha =1.$ See Eq.(1). In this way, we have shown that the inflationary
parameters depend only on the e-folding number $N$ and the final value of
the slow roll parameter $\varepsilon _{end}$. We have also shown that, for
small and negligible running of the scalar spectral index $\frac{dn_{s}}{%
d\ln k}$, the inflationary parameters are in good agreement with WMAP7 data.
On the other hand, we have analyzed the limit $A\longrightarrow 0,$ where
the parameters of inflation depend only on $N$ and the compatibility with
the observations remains conditioned by the values of the number of
e-folding $N$.


\begin{thebibliography}{99}
\bibitem{Braneworld} P.Brax, C.Bruck and A.Davis, $^{\prime \prime }$\textrm{%
Brane World Cosmology,}\emph{\ }$^{\prime \prime }$ \emph{Rept. Prog. Phys.
67 (2004) 2183-2232.}

\bibitem{RSII-1} L. Randall and R. Sundrum, $^{\prime \prime }$\textrm{A
Large Mass Hierarchy from a Small Extra Dimension,}$^{\prime \prime }$ \emph{%
Phys. Rev. Lett. 83 (1999). 3370-3373 ; }L. Randall and R. Sundrum, $%
^{\prime \prime }$\textrm{An Alternative to Compactification,}$^{\prime
\prime }$ \emph{Phys. Rev. Lett. 83 \emph{(1999)} 4690-4693.}

\bibitem{models} R.Maartens, D.Wands, B.Basset and I.Heard, $^{\prime \prime
}$\textrm{Chaotic inflation on the brane,}$^{\prime \prime }$\emph{\ Phys.
Rev. D 62 (2000) 041301}.

\bibitem{BSCZ} A. R. Liddle and A. J. Smith, "\textrm{Observational
constraints on braneworld chaotic inflation,}" \emph{Phys. Rev. D 68 (2003)
061301 }; M. Bennai, H.Chakir and Z.Sakhi, $^{\prime \prime }$\textrm{On
Inflation Potentials in Randall-Sundrum Braneworld Model,}$^{\prime \prime }$
\emph{Electronic Journal of Theoretical Physics 9 (2006) 84--93; }R.
Zarrouki, Z. Sakhi and M. Bennai, $^{\prime \prime }$\textrm{WMAP5
Observationnal Constraints on Braneworld New Inflation Model,}$^{\prime
\prime }$\ \emph{Int. J. Mod. Phys. A 25 (2010) 171-183}.

\bibitem{Chaplygin} R. Jackiw and A.P. Polychronakos, $^{\prime \prime }$%
\textrm{Fluid Dynamical Profiles and Constants of Motion from d-Brane,}$%
^{\prime \prime }$\emph{Commun. Math. Phys. 207 (1999) 107-129.}

\bibitem{DARK M E} DN Spergel, PJ Steinhardt, "\textrm{Observational
evidence for self-interacting cold dark matter,}" \emph{Phys.Rev.Lett.84
(2000) 3760-3763} ; Edmund J. Copeland, M. Sami, Shinji Tsujikawa, "\textrm{%
Dynamics of dark energy}," \emph{Int.J.Mod.Phys.D15(2006)1753-1936}

\bibitem{Herrera1} Ramon Herrera, $^{\prime \prime }$\textrm{Chaplygin
inflation on the Brane,}$^{\prime \prime }$ \emph{Phys. Lett. B 664
(2008)149-153.}

\bibitem{Herrera} Ramon Herrera,$^{\prime \prime }$\textrm{Tachyon-Chaplygin
inflation on the Brane,}$^{\prime \prime }$ \emph{Gen. Rel. Grav. 41 (2009)
1259-1271}

\bibitem{GCG} Mariam Bouhmadi-Lopez, Paulo Vargas Moniz, $^{\prime \prime }$%
\textrm{FRW Quantum Cosmology with a Generalized Chaplygin Gas}$^{\prime
\prime }$ \emph{Phys. Rev. D71 (2005) 063521; }Mubasher Jamil,$^{\prime
\prime }$\textrm{Interacting new generalized Chaplygin gas,}$^{\prime \prime
}$ \emph{Int. J. Theor. Phys. 49 (2010) 62-71; }Mubasher Jamil, $^{\prime
\prime }$\textrm{A single model of interacting dark energy: generalized
phantom energy or generalized Chaplygin gas,}$^{\prime \prime }$ \emph{Int.
J. Theor. Phys. 49 (2010) 144-151.}

\bibitem{Bouhmadi1} Mariam Bouhmadi-Lopez, and Ruth Lazkoz, $^{\prime \prime
}$\textrm{Chaplygin DGP cosmologies}$^{\prime \prime }$ \emph{Phys. Lett. B
654 (2007) 51-57.}

\bibitem{Bouhmadi2} Mariam Bouhmadi-Lopez, Pedro Frazao, and Alfredo B.
Henriques, $^{\prime \prime }$\textrm{Stochastic gravitational waves from a
new type of modified Chaplygin gas,}$^{\prime \prime }$ \emph{Phys. Rev. D
81 (2010) 063504.}

\bibitem{Bertolami} O. Bertolami and V. Duvvuri,$^{\prime \prime }$\textrm{%
Chaplygin Inspired Inflation,}$^{\prime \prime }$ \emph{Phys. Lett. B 640
(2006) 121-125.\ }

\bibitem{Bento} M. C. Bento, O. Bertolami and A. Sen, $^{\prime \prime }$%
\textrm{Generalized Chaplygin Gas, Accelerated Expansion and Dark
Energy-Matter Unification}$^{\prime \prime }$\emph{Phys. Rev. D 66 (2002)
043507.}

\bibitem{Oliver} Oliver F. Piattella, $^{\prime \prime }$\textrm{The Extreme
Limit of the Generalized Chaplygin Gas}$^{\prime \prime }$ \emph{JCAP 03
(2010) 012.. }

\bibitem{Maartens} R. Maartens, $^{\prime \prime }$\textrm{Brane-world
gravity,}$^{\prime \prime }$ \emph{Living Rev. Rel. 7 (2004) 7.}

\bibitem{Lyth} David H. Lyth, Antonio Riotto, "\textrm{Particle Physics
Models of Inflation and the Cosmological Density Perturbation}," \emph{%
Phys.Rept.314(1999)1-146}

\bibitem{Langlois} Langlois, D.; Maartens, R.; Wands, D. "\textrm{%
Gravitational waves from inflation on the brane}," \emph{Phys. Lett. B 489
(2000)259-267. }

\bibitem{Copeland} E.J. Copeland, A.R. Liddle, J.E. Lidsey, $^{\prime \prime
}$\textrm{Steep inflation: Ending braneworld inflation by gravitational
particle production,}$^{\prime \prime }$ \emph{Phys. Rev. D 64 (2001) 023509.%
}

\bibitem{Sami} M. Sami, P. Chingangbam, T. Qureshi, $^{\prime \prime }$%
\textrm{Aspects of Tachyonic Inflation with Exponential Potential,}$^{\prime
\prime }$ \emph{Phys. Rev. D 66 (2002) 043530.}

\bibitem{WMAP7} E. Komatsu et al.,$^{\prime \prime }$\textrm{Seven year
Wilkinson Microwave Anisotropy Probe (WMAP) observations: power spectra and
WMAP-deriverd\ parameters}$^{\prime \prime }\ $\emph{arXiv}$:$ \emph{%
1001.4635\ v1, }Astrophysical Journal Supplement Series\ (2010).
\end{thebibliography}
\end{document}